\documentclass{aa}

\usepackage[applemac]{inputenc} 

\usepackage{subfig}
\usepackage{natbib}  
\bibpunct{(}{)}{;}{a}{}{,} 
\usepackage{array} 
\usepackage{tabularx}
\usepackage{graphicx}
\usepackage[version=3]{mhchem}
\usepackage{txfonts}

\begin{document}

\title{High-temperature measurements of VUV-absorption cross sections of \ce{CO2} and their application to exoplanets}
\author{O. Venot\inst{\ref{LAB},} \inst{\ref{CNRS}}
\and N. Fray\inst{\ref{LISA},}
\and Y. B\'{e}nilan\inst{\ref{LISA},}
\and M.-C. Gazeau\inst{\ref{LISA},}
\and E. H\'{e}brard\inst{\ref{LAB},}\inst{\ref{CNRS}}
\and G. Larcher\inst{\ref{LISA},}
\and M. Schwell\inst{\ref{LISA},}
\and M. Dobrijevic\inst{\ref{LAB},}\inst{\ref{CNRS}}
\and F. Selsis\inst{\ref{LAB},}\inst{\ref{CNRS}}}

\institute{Univ. Bordeaux, LAB, UMR 5804, F-33270, Floirac, France\\\email{venot@obs.u-bordeaux1.fr}\label{LAB}
\and CNRS, LAB, UMR 5804, F-33270, Floirac, France\label{CNRS}
\and Laboratoire Interuniversitaire des Syst\`{e}mes Atmosph\'{e}riques, UMR CNRS 7583, Universit\'{e}s Paris Est Cr\'eteil (UPEC) et Paris Diderot (UPD), Cr\'{e}teil, France\label{LISA}}

\date{Received <date> /
Accepted <date>}

\abstract{
Ultraviolet (UV) absorption cross sections are an essential ingredient of photochemical atmosphere models. Exoplanet searches have unveiled a large population of short-period objects with hot atmospheres, very different from what we find in our solar system. Transiting exoplanets whose atmospheres can now be studied by transit spectroscopy receive extremely strong UV fluxes and have typical temperatures ranging from 400 to 2500~K. At these temperatures, UV photolysis cross section data are severely lacking.}
{
Our goal is to provide high-temperature absorption cross sections and their temperature dependency for important atmospheric compounds. This study is dedicated to \ce{CO2}, which is observed and photodissociated in exoplanet atmospheres. We also investigate the influence of these new data on the photochemistry of some exoplanets.}
{
We performed these measurements with synchrotron radiation as a tunable VUV light source for the 115 - 200~nm range at 300, 410, 480, and 550~K. In the 195 - 230~nm range, we used a deuterium lamp and a 1.5 m Jobin-Yvon spectrometer and we worked at seven temperatures between 465 and 800~K. We implemented the measured cross section into a 1D photochemical model.}
{
For $\lambda >$ 170~nm, the wavelength dependence of $\ln(\sigma_{\ce{CO2}}(\lambda, T) \times \frac{1}{Q_v(T)})$ can be parametrized with a linear law. Thus, we can interpolate $\sigma_{\ce{CO2}}(\lambda, T)$ at  any temperature between 300 and 800 K. Within the studied range of temperature, the \ce{CO2} cross section can vary by more than two orders of magnitude. This, in particular, makes the absorption of \ce{CO2} significant up to wavelengths as high as 230~nm, while it is negligible above 200~nm at 300~K.}
{
The absorption cross section of \ce{CO2} is very sensitive to temperature, especially above 160~nm. The model predicts that accounting for this temperature dependency of \ce{CO2} cross section can affect the computed abundances of \ce{NH3}, \ce{CO2}, and CO by one order of magnitude in the atmospheres of hot Jupiter and hot Neptune. This effect will be more important in hot \ce{CO2}-dominated atmospheres.}

\maketitle

\section{Introduction}

Exoplanets exhibit a wide variety of mass, radius, orbits, and host stars. Because of observational biases, most transiting exoplanets are very close to their parent stars and are highly irradiated, implying large UV fluxes and high atmospheric temperatures. The atmosphere of transiting hot Jupiters and hot Neptunes can be studied by spectroscopy at the primary transit \citep{tinetti2007water, tinetti2007infrared, swain2008presence, beaulieu2010water, Tinetti2010, 2011beaulieu} and at the secondary eclipse \citep{swain2009water, swain2009molecular, stevenson2010possible, stevenson2012two}. Photochemistry has an important influence on the atmospheric composition of these exoplanets, from the top of the atmosphere down to 100~mbar \citep{moses2011disequilibrium, line2011thermochemical, venot2012}. For these exoplanets and within this large pressure range, the temperature can vary roughly from 400 to 2500~K. To model correctly the photochemistry of these planets, we need to use absorption cross sections consistent with these temperatures for all the species whose photolysis plays an important role in either the formation/destruction of molecules or in the penetration of the UV flux into the atmosphere. Carbon dioxide (\ce{CO2}) is one of these species. It has been observed in extrasolar giant planet atmospheres \citep{swain2009water, swain2009molecular}, but cross section measurements ($\sigma_{\ce{CO2}}(\lambda,T)$) are extremely sparse above room temperature. 

Some high-temperature measurements have been performed in the past but only at a few wavelengths. For instance, \cite{Koshi1991519} measured the production of O($^3$P) in the photodissociation of \ce{CO2} at 193~nm by using atomic resonance absorption spectroscopy behind reflected shock waves between 1500 and 2700~K. They saw that the production of oxygen atoms increases with the temperature. Before that, \cite{generalov1963absorption} measured $\sigma_{\ce{CO2}}(\lambda,T)$ behind a shock wave at temperatures up to 6300~K at 238 and 300~nm, and observed absorption up to 355~nm at 5000~K. These measurements, because of the technique employed, are limited to a narrow range of wavelengths and do not provide complete spectra. Nevertheless, they showed that the absorption of \ce{CO2} increases at high temperature.

The first experiments dedicated to the determination of absorption cross sections of \ce{CO2} at temperatures different from 298~K were motivated by solar system planetary studies (Mars, Titan, Venus, primitive Earth). \cite{Lewis1983297} measured $\sigma_{\ce{CO2}}(\lambda,T)$ between 120 and 197~nm at 202 and 367~K. They observed an enhancement of $\sigma_{\ce{CO2}}(\lambda,T)$ for the longer wavelengths when the temperature increases. This trend was confirmed by \cite{yoshino96b}, who measured the absorption cross section of \ce{CO2} at 195 and 295~K, between 118.7 and 175.5~nm. \cite{parkinson2003} extended these measurements up to 192.5~nm for 195~K, and up to 200 nm for 295~K. They observed that the cross sections at 195~K were smaller than those at 295~K. Finally, still in the frame of Mars and Venus studies, \cite{stark2007} explored a lower wavelength range and measured $\sigma_{\ce{CO2}}(\lambda,T)$ between 106.1 and 118.7~nm at 295 and 195~K.

During the same period, \cite{jensen1997ultraviolet} used a heated cell traversed by a tunable laser to measure the absorption cross section of \ce{CO2} at high temperatures. They obtained spectra from 230 to 355~nm at 1523, 1818, 2073, and 2273~K. In the frame of combustion studies, \cite{Schulz200282} measured the absorption cross section in shock-heated \ce{CO2} between 190 and 320~nm, and for temperatures ranging from 900 to 3050~K. They fit the strong temperature dependence of $\sigma_{\ce{CO2}}(\lambda,T)$ with an empirical function. \cite{oehlschlaeger2004ultraviolet} extended these data to higher temperatures (up to 4500~K). They measured the absorption of shock-heated carbon dioxide at four different laser wavelengths (216.5, 244, 266, and 306~nm). They also fit the variation of $\sigma_{\ce{CO2}}(\lambda,T)$ with a semi-empirical formula. \cite{jeffries2005uv} showed that the temperature dependence of \ce{CO2} absorption in the UV could be used to determine the gas temperature, making this parameter a useful tool in combustion applications.

To our best knowledge, no measurements exist of the absorption cross section of \ce{CO2} between 300 and 900~K in the wavelength range useful for exoplanetary studies ($<$190~nm). Here we report measurements of this parameter at 300, 410, 480, 550~K between 115 and 200~nm, as well as between 195 and 230~nm at seven temperature values between 465 and 800~K. We determine a semi-empirical formula to fit the temperature dependence for wavelengths longer than 170~nm. Finally, we study the effect of these new data on the atmospheric composition predicted by a 1D photochemical model of hot exoplanet atmospheres.

\begin{figure*}[!hbt]
\centering
\includegraphics[width=0.90\textwidth]{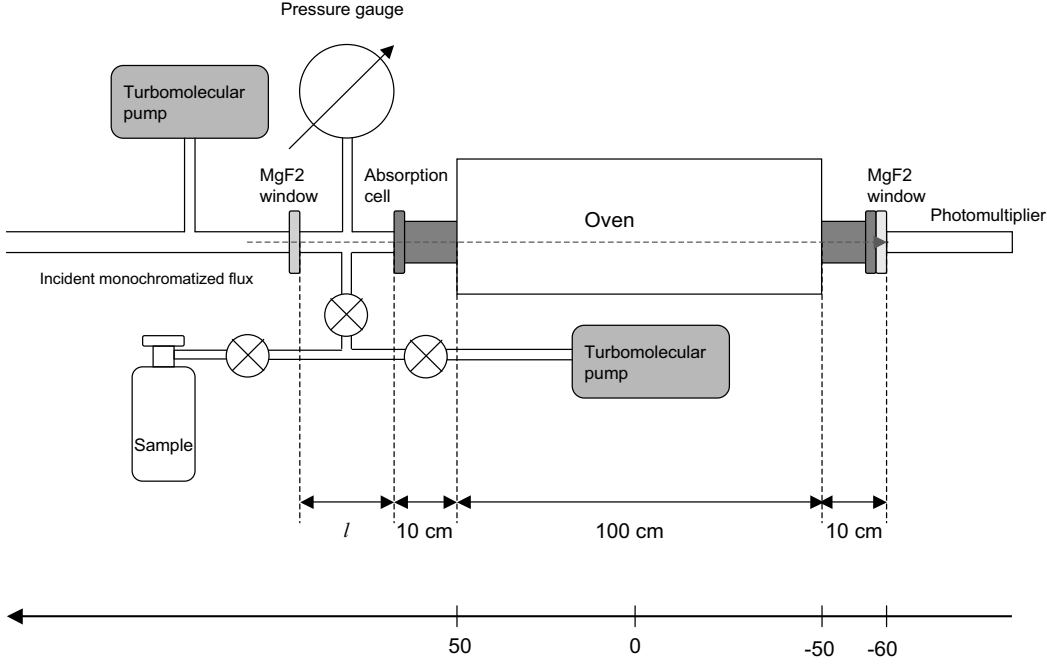}
\caption{Experimental setup used for measuring the ultraviolet absorption spectra at high temperatures. The length $l$ changed during the different set of measurements. For the 115-200~nm experiments, $l$ = 13.3 cm. For the 195-230~nm range, $l$ = 27.0~cm. Finally, for the temperature calibration, $l$ = 20~cm.}
\label{fig:schema}
\end{figure*}

\section{Experimental Methods and Procedures}\label{sec:procedures}
\subsection{Measurements}

We used gaseous \ce{CO2} of 99.995\% purity.
Tunable VUV light between 115 and 200~nm was obtained from the synchrotron radiation facility BESSY in Berlin. Measurements in this spectral range were performed using a three-meter focal length normal incidence monochromator (NIM) equipped with a 600~lines/mm holographically ruled grating with a linear dispersion of 0.56~nm.mm$^{-1}$ and connected to a dipole magnet beamline (DIP12-1B)\citep{reichardt2001b}. We recorded spectra of \ce{CO2} with a resolution of 0.05~nm. 
Wavelength calibration was obtained by using the set of measurements of \cite{yoshino96b}, \cite{parkinson2003}, and \cite{stark2007} as a reference.
The VUV radiation intensity was measured directly with a solar blind photomultiplier tube closed by a MgF$_2$ window (Electron Tubes Limited 9402B with caesium-telluride photocathode, see Fig. \ref{fig:schema}). The entrance of the cell was also closed by a MgF$_2$ window. With this configuration, the absorption cell is a cylinder with an optical path length of 133~cm. 
The signal coming from the photomultiplier was recorded through a pico ampere meter (Keithley) using an integration time of 1s per point. We recorded three points per resolution interval.

To account for the steady decrease of the incident VUV light intensity, which is caused by the decay of BESSY's storage ring current, two "empty-cell" spectra were recorded just before and right after each \ce{CO2} spectrum. A synthetic empty-cell spectrum was then interpolated by considering a linear decrease of the incident light intensity in accordance with the recorded ring current decrease during our experiment time (1/2 hour approximately). Comparing the spectra we have acquired at different pressures, we estimate that this procedure leads to an uncertainty of 10\% on the absorption cross section and that this error dominates over all other sources of errors.

Measurements in the 195 - 230~nm range were performed at the Laboratoire Interuniversitaire des Syst\`{e}mes Atmosph\'{e}riques (LISA) in Cr\'{e}teil, France. We used a UV deuterium lamp and a 1.5~m Jobin-Yvon spectrometer.  The experimental setup was the same as for the lower wavelength range (see Fig. \ref{fig:schema}), except that the optical path length was 147~cm. In this spectral range, we recorded spectra with a resolution of 0.3~nm. In both cases, an oven (Nabertherm) was used to heat the cell to a temperature of 1400~K. The temperature of the oven was constant and measured continuously at three fixed points along the tube.


Gaseous carbon dioxide was introduced into the cell at the desired pressure, which was monitored by two MKS Baratron \textregistered manometers (range 10$^{-4}$ to 1 mbar and 1 to 1000 mbar). All spectra were taken for at least three different pressures in the 0.2-1000~mbar range to check the reproducibility of our measurements. Comparing all the spectra we have acquired, we restricted the influence of the stray light by selecting only spectra that were free from any saturated absorption, i.e., with transmission greater than 10\%.

\subsection{Temperature calibration}\label{sec:t_calib}

\begin{figure}[!ht]
\includegraphics[width=1\columnwidth]{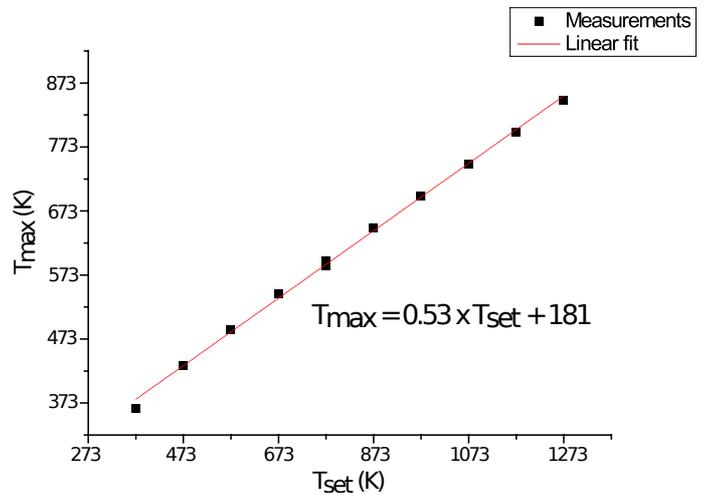}
\caption{Relation between the set temperature ($T_{set}$) and the temperature measured at the center of the oven ($T_{max}$).}
\label{fig:temp_meas}
\end{figure}

Even if the temperature of the oven is controlled at three different points, the temperature distribution of the gas inside the cell is not homogeneous along the optical pathway. Moreover, the temperature set by the oven ($T_{set}$) does not correspond to the effective temperature of the gas ($T_{gas}$). We calibrated the temperature of the gas in two steps. First, with a thermocouple type E, we measured the temperature at the center of the absorption cell, without gas. Figure \ref{fig:temp_meas} shows the relation between the set temperature and the temperature measured inside the cell at the center $T_{max}$. As we can see, for $T_{set}$=1273~K, the temperature at the center of the oven is only 856~K. We find that the maximum temperature in the cell is given by

\begin{equation}\label{eq:tmax}
T_{max} = 0.53 \times T_{set} + 181.
\end{equation}

\begin{figure}
\includegraphics[width=\columnwidth]{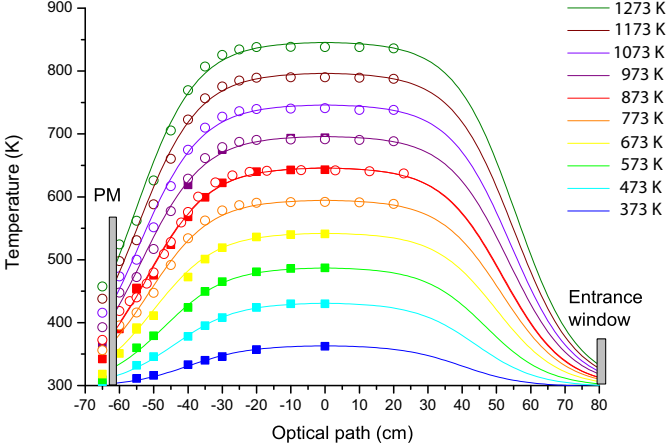}
\caption{Temperature gradient inside the cell. 0~cm is the middle of the oven, where the temperature is the highest ($T_{max}$).}
\label{fig:grad_temp}
\end{figure}

\begin{figure}[!ht]
\includegraphics[width=1\columnwidth]{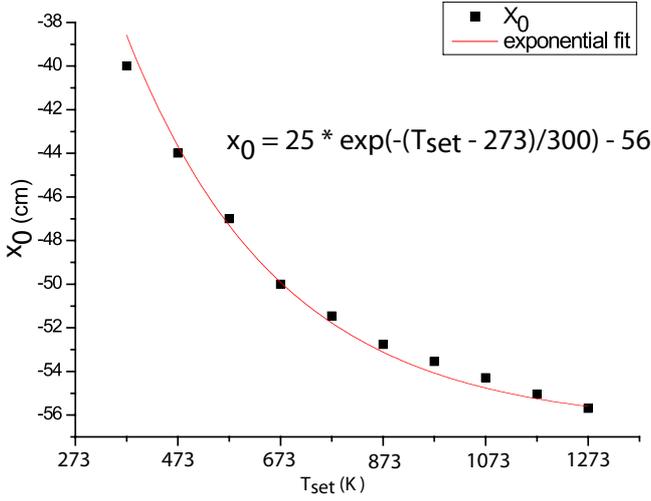}
\caption{Relation between $x_0$ and the order temperature $T_{set}$.}
\label{fig:x0_Tset}
\end{figure}

We then measured the temperature gradient inside the absorption cell at different points in the cell (see Fig.~\ref{fig:grad_temp}). Figure~\ref{fig:grad_temp} also shows that there is a middle area where the temperature is relatively constant, and the temperature decreases towards the extremities of the absorption cell. The temperature in the cell $T_{gas}(x)$ can be modeled using a symmetrized inverse exponential function with two limits

\begin{equation}\label{eq:Tx}
T_{gas}(x)=T_{max} + \frac{T_{amb} - T_{max}}{1+exp(-\frac{|x|-|x_0|}{\Delta x})},
\end{equation}
where $T_{max}$ (K) is calculated from Eq.~\ref{eq:tmax}, and $T_{amb}$ is the ambient temperature (298~K); $x_0$ (cm) and $\Delta x$ (cm) are determined by minimizing the $\chi^2$ function using the measured data. $x_0$ corresponds to the position where $T_{gas}(x_0) = (T_{amb}+T_{max})/2$. The cell is heated by the oven and cooled by conduction from the part at ambient temperature outside the oven. An equilibrium is reached by the two processes that fix $x_0$. Figure~\ref{fig:x0_Tset} shows the relation between $x_0$ and $T_{set}$. We find that $x_0$ is given by

\begin{equation}
x_0=25 \times \exp(-\frac{T_{set}-273}{300})-56
\end{equation}
with $x_0$ in cm and $T_{set}$ in Kelvin.
As we can see in Fig.~\ref{fig:x0_Tset}, $|x_0|$ varies by 30\%, from 40 cm to 56 cm, in our temperature range. The fact that $x_0$ increases with temperature means that the cooling process becomes less efficient when the temperature increases. The value of $\Delta x$ is found to be constant for the whole temperature range and is equal to 9 cm. 

\subsection{Calculation of the absorption cross section}

When the temperature is constant in the cell, absolute photoabsorption cross sections can be calculated using the Beer-Lambert law
\begin{equation}\label{eq:sigmaTconst}
\sigma = \left(\frac{1}{nL}\right) \times \ln \left(\frac{I_0}{I}\right),
\end{equation}
where $\sigma$ corresponds to the absorption cross section (cm$^2$), $I_0$ is the light intensity transmitted with an empty cell, $I$ is the light intensity transmitted through the gas sample, $L$ is the absorption path length (cm), and $n$ is the volume density of the gas (cm$^{-3}$), following the relation $P=nk_BT$, where $T$ (K) and $P$ (Pa) are, respectively, the temperature and the pressure of the sample and $k_B$ the Boltzmann constant. 

\begin{figure}[!ht]
\includegraphics[width=\columnwidth]{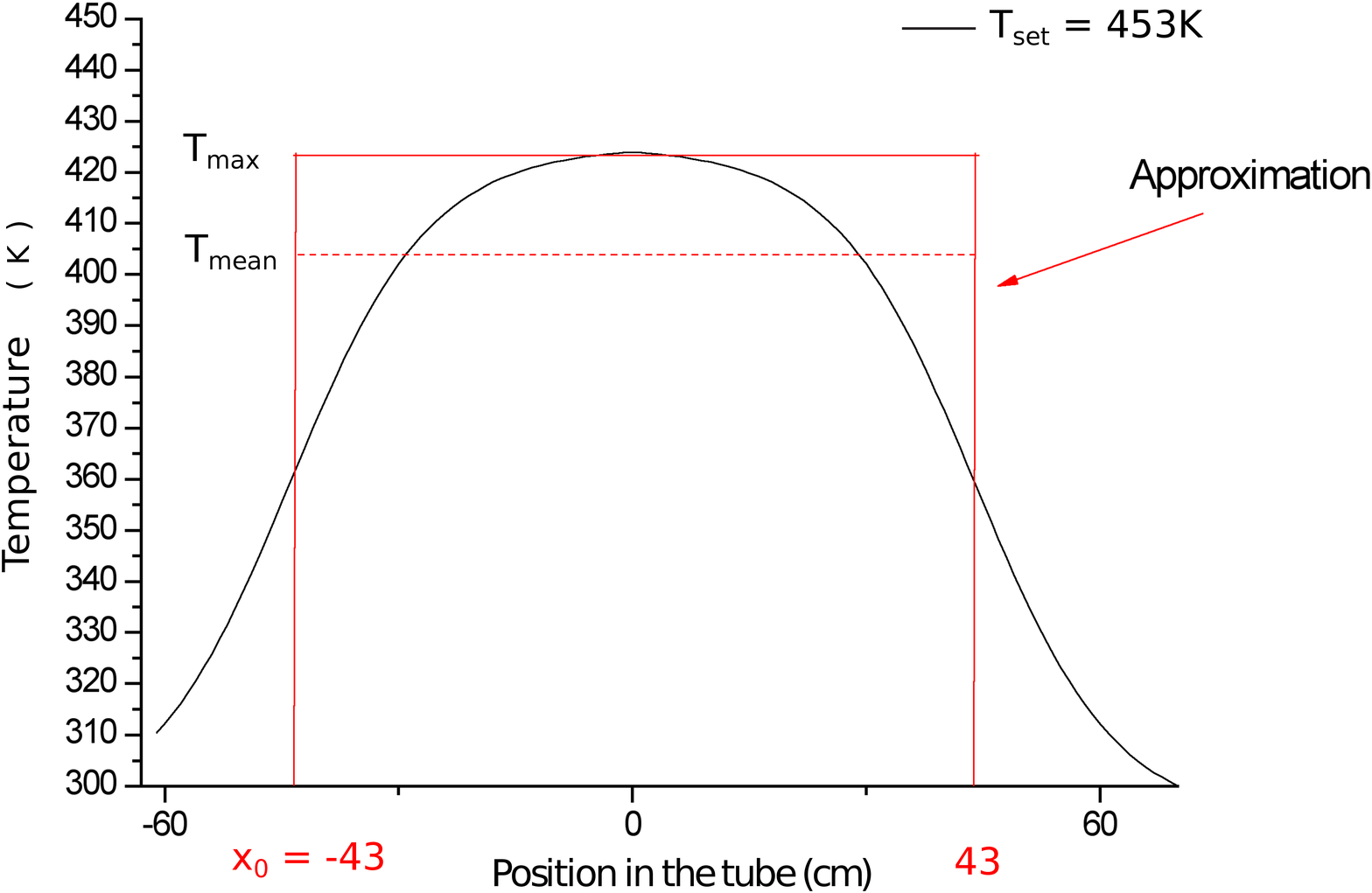}
\caption{Temperature inside the absorption cell for $T_{set}=453~K$. The maximum temperature $T_{max}$ is 420~K and the mean temperature $T_{mean}$ is (410$\pm$15)~K. The portion of gas out of the range [$-x_0 ; x_0$] is considered at ambient temperature.}
\label{fig:Tmean180}
\end{figure}

As explained in Sect.~\ref{sec:t_calib}, in our experiments the temperature is not constant along the optical pathway, so we can not use the Eq.~\ref{eq:sigmaTconst} since the density depends on $T$. Consequently, the light transmitted through the cell should be integrated step by step along the optical pathway and the data should be inverted in order to retrieve the absorption cross sections. In order to simplify the problem, we consider a mean temperature $T_{mean}$ for the gas along the pathway $-x_0$ to $x_0$ (Fig.~\ref{fig:Tmean180}). This way, we can assign one spectrum to one temperature. The mean temperature is calculated by integrating Eq.~\ref{eq:Tx} from $-x_0$ to $x_0$ and normalizing by 2$x_0$. The portion of gas out of the range [$-x_0 ; x_0$] is considered to be at ambient temperature. The mean temperature calculated is then used to calculate the absorption cross section. For instance, with $T_{set}$= 453~K, the maximum temperature reached is in fact $T_{max}$=420~K and we determined an absorption cross section for a mean temperature $T_{mean}$= (410$\pm$15)~K. In our case, the intensity $I$ is attributable partly to the section at the ambient temperature and partly to the section at the mean temperature.

\begin{figure}
\includegraphics[width=1\columnwidth]{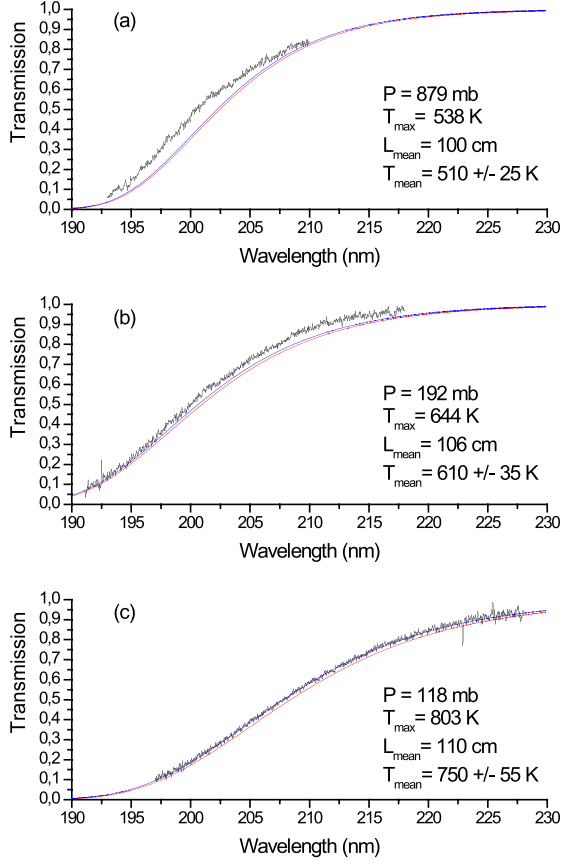}
\caption{Transmission of the flux as a function of the wavelength determined by the measurements (black lines) compared to the measurements obtained from the complete radiative transfer through the cell, taking into account the real temperature variation along the optical path, (a) 538~K, (b) 644~K, and (c) 803~K (red line), and the measurements obtained when considering the mean temperature (a) 510~K, (b) 610~K, and (c) 750~K (blue line).}
\label{fig:transmission}
\end{figure}

The absorbance of the portion of gas at ambient temperature is subtracted from the overall absorbance measured with the heated gas. We calculate the absorption cross section at the mean temperature with the formula
\begin{equation}
\sigma(T_{mean}) = \frac{1}{ n_{mean} \times L_{mean}} \times  \left[ \ln \left(\frac{I_0}{I}\right) - \sigma(T_{amb}) \times n_{amb} \times L_{amb}\right],
\end{equation}
where $\sigma(T_{amb})$ and $\sigma(T_{mean})$ (cm$^2$) are the absorption cross sections at ambient temperature and at the mean temperature of the gas, $n_{amb}$ and $n_{mean}$ (cm$^{-3}$) are the volume densities of the gas in the portion at $T_{amb}$ and at $T_{mean}$, respectively. $L_{mean} = 2 x_0$ and $L_{amb} = L-2x_0$ (cm) are the portion of gas at $T_{mean}$ and at $T_{amb}$, respectively. We validated a posteriori this approximation by comparing measured transmissions at different temperatures and pressures, with the transmissions obtained from the complete radiative transfer through the cell taking into account the temperature variation along the optical path and the temperature variation of the absorption cross section presented in the next section. Results are generally in agreement to better than one percent (see Fig.~\ref{fig:transmission}).

\section{Results and Discussion}\label{sec:results}

\subsection{Photoabsorption cross section from 115 nm to 200 nm}

Before heating the gas, we measured ambient temperature (300~K) spectra of \ce{CO2} in order to calibrate and compare it with the previously published data \citep{yoshino96b, parkinson2003, stark2007}; our measurements agree very closely with these measurements with a difference of less than a few percentage points for all wavelengths. Our measurements at room temperature did not go up to 200 nm, so between 170 and 200 nm we use the data of \cite{parkinson2003}.\\

\begin{figure}
\includegraphics[width=1\columnwidth]{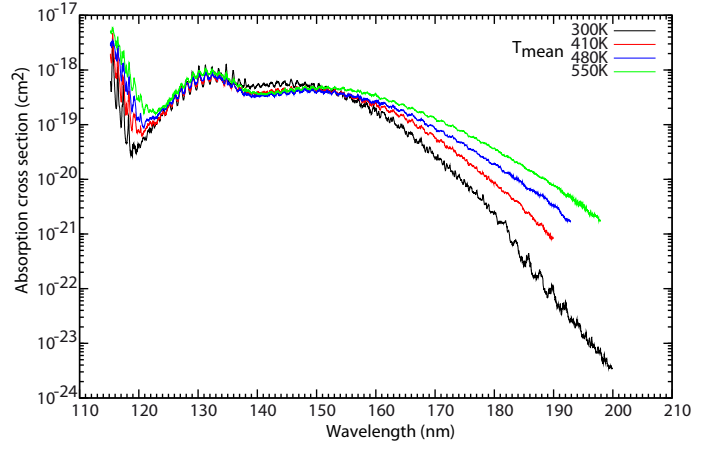}
\caption{Absorption cross section of \ce{CO2} at $T_{mean}$ = 300~K (black), 410~K (red), 480~K (green) and 550~K (blue) for wavelengths between 115 and 200~nm.}
\label{fig:BESSY}
\end{figure}

Then, we measured $\sigma_{\ce{CO2}}(\lambda, T)$ at three different temperatures: 410 ($\pm$~15) K, 480 ($\pm$~25) K, and 550 ($\pm$~30) K. We show these data in Fig.~\ref{fig:BESSY}.
Between 115 and 120~nm we see a change of the cross section which depends on the temperature. At 120~nm, the absorption cross section is ten times higher at 550~K than at 300~K and an increase of a factor of 2.5 is observed at 121.6 nm between the lower and the upper temperature. Slight differences of up to 50\% can be observed between 125 and 140~nm while between 140 and 150~nm differences are minor.
After 160~nm, we clearly observe large differences between the different temperatures. The slope of the cross section varies with the temperature. The higher the temperature is, the less steep is the slope. At 195~nm, there is a factor $\sim$200 between $\sigma_{\ce{CO2}}$($\lambda$,~300~K) and $\sigma_{\ce{CO2}}$($\lambda$,~550~K).\\

\subsection{Photoabsorption cross section from 195 nm to 230 nm}

We measured $\sigma_{\ce{CO2}}(\lambda, T)$ at seven different temperatures: 465~($\pm$~20)~K, 510~($\pm$~25)~K, 560~($\pm$~30)~K, 610~($\pm$~35)~K, 655~($\pm$~45)~K, 750~($\pm$~55)~K and 800~($\pm$~60)~K. As for the cross section at shorter wavelengths, we clearly see the dependence on the temperature in this wavelength range and the increase of the cross section for high temperatures (Fig.~\ref{fig:PARAM}). As we plotted the data obtained previously at shorter wavelengths in this figure, we see good agreement between the two ranges. Especially, we see that $\sigma_{\ce{CO2}}$($\lambda <$ 200~nm, 550~K) matches almost perfectly with $\sigma_{\ce{CO2}}$($\lambda >$ 195~nm, 560~K).

\subsection{Determination of an empirical law}

\begin{figure}[!ht]
\includegraphics[width=1\columnwidth]{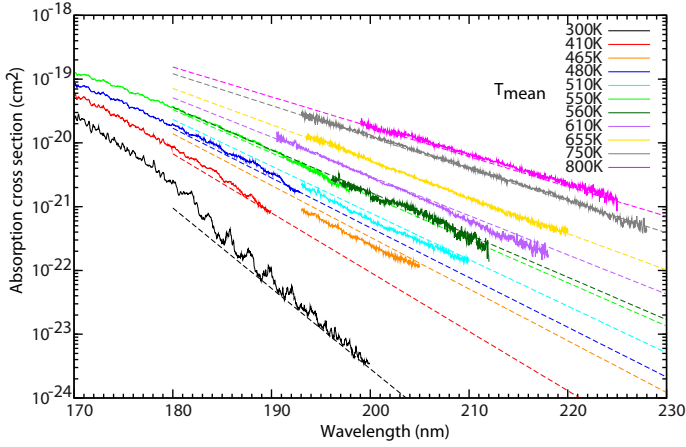}
\caption{Absorption cross section of \ce{CO2} for wavelengths longer than 195~nm at 465~K, 510~K, 560~K, 610~K, 655 K, 750~K, and 800~K, plotted with the cross section at ambient temperature (black) and the absorption cross sections measured at shorter wavelengths and presented in Fig.~\ref{fig:BESSY} (300~K, 410~K, 480~K, and 550~K). The absorption cross sections calculated with Eq.~\ref{eq:formule} are plotted with the same color coding.}
\label{fig:PARAM}
\end{figure}

For wavelengths longer than 170 nm, we parametrize the variation of $\ln(\sigma_{\ce{CO2}}(\lambda,T) \times \frac{1}{Q_v(T)})$ with a linear regression

\begin{equation}\label{eq:formule}
\ln \left(\sigma_{\ce{CO2}}(\lambda,T) \times \frac{1}{Q_v(T)}\right) = a(T) + b(T) \times \lambda
\end{equation}
with $T$ in K and $\lambda$ in nm and where \\
$a(T)= -42.26 + (9593\times 1.44/T)$, \\
$b(T) =  4.82\times10^{-3} - 61.5\times 1.44/T$,\\
and \\
$Q_v (T) = (1- \exp(-667.4\times 1.44/T))^{-2} \times (1-\exp(-1388.2 \times 1.44/T))^{-1} \times (1-\exp(-2449.1\times 1.44/T))^{-1}$\\
is the partition function. Figure \ref{fig:PARAM} compares the absorption cross sections obtained with this calculation to the measurements. We can see that the parametrization is very good.

In this wavelength region, measurements at ambient temperature have been obtained with high resolution \citep{cossart1992high, cossart2005high}. The narrow bands that have been identified in this study allow us to make the assumption that the continuum is made of the superposition of bands corresponding to transitions from high vibrational states of the electronic ground state to different vibrational states of the upper electronic states $^1$B$_2$, $^1$A$_2$, or $^3$B$_2$ \citep[see][]{cossart2005high}. If we suppose that at each wavelength we associate only one transition coming from a defined vibrational level having an effective energy, the cross sections which are proportional to the population of the lower level should have a Boltzmann like temperature dependence ($\ln \sigma_{\ce{CO2}}(\lambda,T) \times Q_v(T) = -hc\nu/k_BT+$Cste). So, the coefficients $a(T)$ and $b(T)$ plotted as a function of $hc/k_BT(=1.44/T)$ should give a straight line from which we could obtain the ground state frequency of the vibration mode involved in the electronic transition. Trying to use this model did not give a good parametrization of our data. Consequently, we chose to keep our parametrization given by Eq.~\ref{eq:formule} which has no physical meaning but simply allows us to parametrize our data with only four parameters. This greatly simplifies the implementation of the temperature dependency of the absorption cross sections in the radiative transfer codes. 

%

Our results are not compatible with the data of \cite{Schulz200282}. We used their formulation $\ln \sigma(\lambda, T) = a + b \lambda$ with the coefficients $a$ and $b$ given in their paper, to determine the absorption cross section of \ce{CO2} for some temperatures in the range 900-3500 K. The absorption cross section calculated for $T$=940~K falls in between our data for 655~K and 750~K. Also, the value found with their calculation for $T$=1160~K is only slightly higher than our data at 800~K. \cite{Schulz200282} may have overestimated the temperature in their measurements. Or, another possible problem in their measurement is that their background signal ($I_0$) is measured at ambient temperature. Thermal emission is therefore not taken into account when their samples are heated. This could lead to an underestimation of their absorbance, and consequently to lower absorption cross sections compared to ours.

\section{Application to exoplanets}\label{sec:application}

\begin{figure}[!ht]
\includegraphics[width=1\columnwidth]{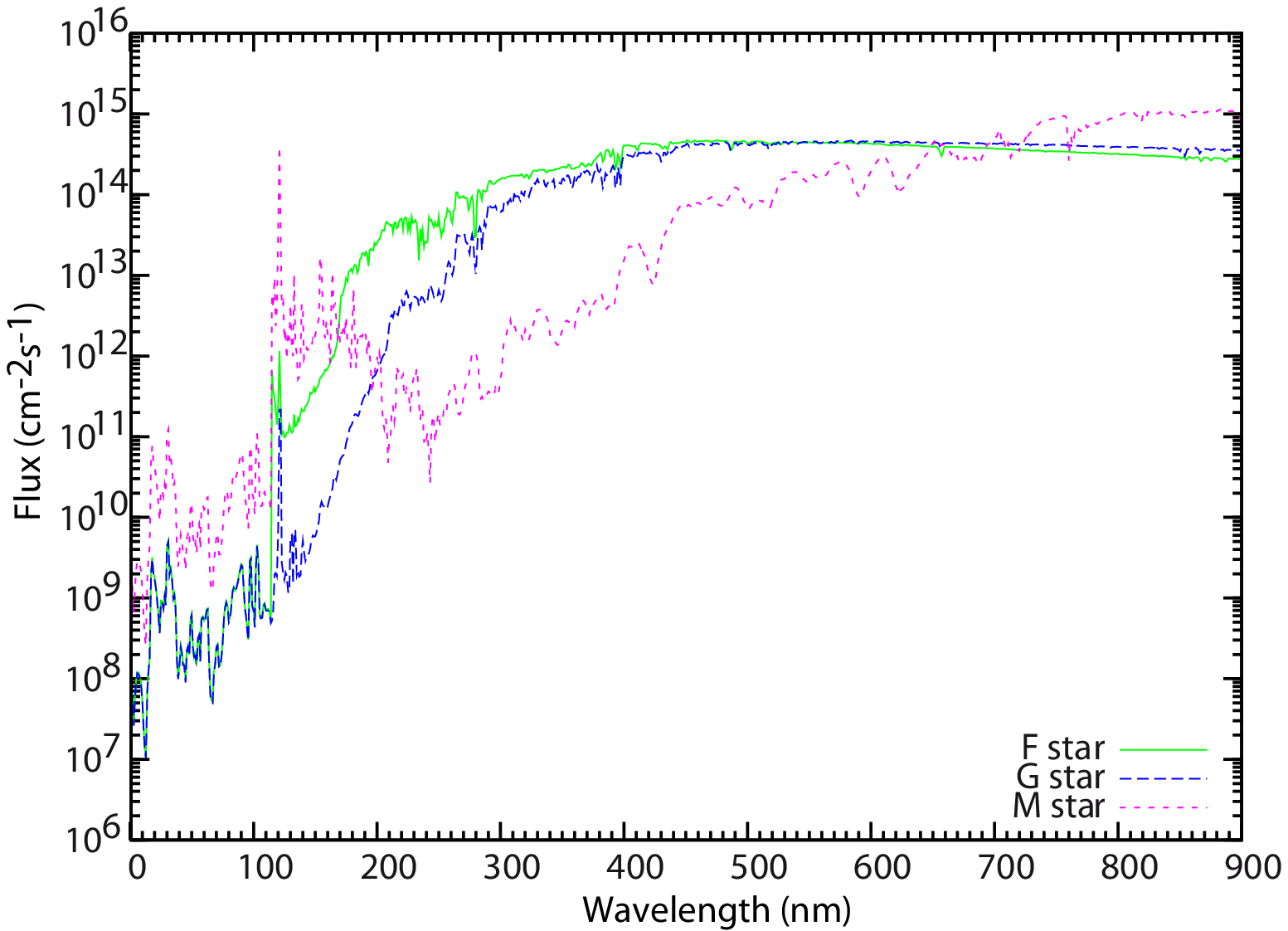}
\caption{UV flux received at the top of the atmosphere of a GJ~436b-like planet with the M star (dotted line pink), G star (dashed line blue) and F star (full line green).}
\label{fig:flux_top}
\end{figure}

We investigated the impact of the temperature dependency of the \ce{CO2} cross section on a prototype planet whose characteristics are similar to those of the hot Neptune GJ 436b. We chose this planet because the temperature of its upper atmosphere is around $\sim$500~K which corresponds to the highest temperature for which we measured the cross section between 115 and 200 nm. This exoplanet was discovered in 2004 by \citet{butler2004neptune}. Its semi-major axis is $a$ = 0.02887 ($\pm$ 0.00095) AU and its mass is M.sin$i$ = 0.0737 ($\pm$ 0.0052) M$_J$ \citep{southworth2010homogeneous}.We considered three different types of host stars: an M, a G, and an F star. We used the spectra of GJ 644 (M3V, \citealt{segura2005biosignatures}), the Sun (G2V, \citealt{gueymard2004}), and HD 128167 (F2V, \citealt{segura2003ozone}), scaled to get the bolometric flux received by GJ436b. Figure~\ref{fig:flux_top} compares the flux received at the top of the atmosphere of a GJ~436b-like planet with the different stars.

We used the model described in \cite{venot2012} and the same temperature profile as \cite{line2011thermochemical} calculated by \cite{lewis2010atmospheric}. Although this is not a realistic assumption, we used the same temperature for all three host stars.
\begin{figure}[!ht]
\includegraphics[width=1\columnwidth]{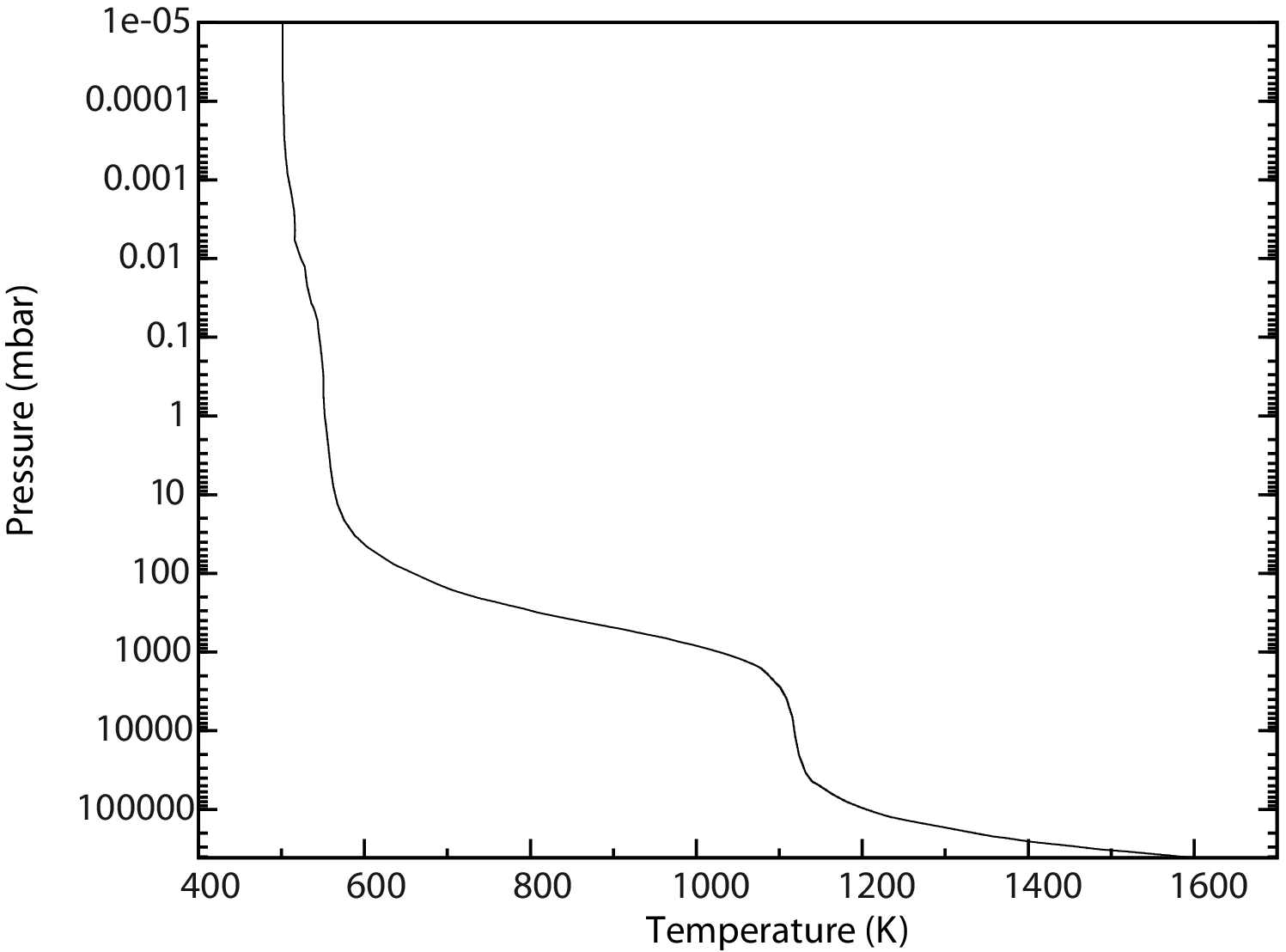}
\caption{Temperature profile of GJ~436b \citep{lewis2010atmospheric}.}
\label{fig:profilPT}
\end{figure}
To model the vertical mixing, we considered an eddy diffusion coefficient constant $K_{zz}=10^{8}$cm$^2$.s$^{-1}$. Elemental abundances of the atmosphere of this planet are highly uncertain \citep{stevenson2010possible, madhusudhan2011highGJ436b}. So we assumed a heavy elemental enrichment of 100 compared with solar abundances \citep{grevesse1998standard}, which is arbitrary but higher only by a factor of 2 than the carbon enrichment of Uranus and Neptune (\citealt{hersant2004enrichment} and references therein). Consequently we obtained a high abundance of \ce{CO2}.


\ce{CO2} has two routes to photodissociate :
\begin{align*}
\ce{CO2} + h\nu \ce{-> CO + O(^3P)} \qquad J4\\
\ce{CO2} + h\nu \ce{-> CO + O(^1D)} \qquad J5
\end{align*}
Depending on the energy of the photons, one route is favored over the other. The quantum yield for these two photolyses, $q_4(\lambda)$ and $q_5(\lambda)$, are presented in Table~\ref{tab:q} \citep{huebner1992solar}. We see that at longer wavelengths, \ce{CO2} photodissociates through the route $J4$. We assume that it remains the same at high temperature.

\begin{table}[!h]
\begin{center}\begin{tabular}{ll}
\hline
\hline
Quantum yield & Values [wavelength range] \\
\hline
$q_4(\lambda)$ & 1 [167-227]  \\
$q_5(\lambda)$ & variable [50-107] ; 1 [108-166] \\
\hline
\end{tabular}\end{center}
\caption{Quantum yields for the photodissociations of \ce{CO2}}\label{tab:q}
\end{table}

The loss rate of a photolysis $Jk$ corresponds to the loss of \ce{CO2} because of this photolysis. It is given by
\begin{equation}
\frac{\partial n_{\ce{CO2}}}{\partial t} = - J^k_{\ce{CO2}} n_{\ce{CO2}},
\end{equation}
where $n_{\ce{CO2}}$ is the density of \ce{CO2} (cm$^{-3}$) and $J^k_{\ce{CO2}}$ is the photodissociation rate of \ce{CO2} by the photodissociation $Jk$

\begin{equation}
J_{\ce{CO2}}^k(z) = \int_{\lambda_1}^{\lambda_2} \sigma_{\ce{CO2}}(\lambda) F(\lambda, z) q_k(\lambda) d\lambda,
\end{equation}
where $[\lambda_1;\lambda_2]$ is the spectral range on which \ce{CO2} absorbs the UV flux, $\sigma_{\ce{CO2}}(\lambda)$ the absorption cross section of \ce{CO2} at the wavelength $\lambda$ (cm$^2$), $F(\lambda, z)$ the UV spectral irradiance at $\lambda$ and the altitude $z$ (cm$^{-2}$.s$^{-1}$.nm$^{-1}$), and $q_k(\lambda)$ the quantum yield corresponding to the photodissociation $Jk$. \\

First, we find the steady-state composition of these atmospheres using the absorption cross sections available in the literature, which means at ambient temperature. Then, we replace the "ambient cross section" of \ce{CO2} ($\sigma_{\ce{CO2}}$(300~K)) by the cross section measured at 550~K, between 115 and 200 nm($\sigma_{\ce{CO2}}$(550~K)). For wavelengths between 200 and 230 nm, we use Eq.\ref{eq:formule} to determine $\sigma_{\ce{CO2}}$(550~K).\\

\begin{figure*}[!ht]
\includegraphics[height=0.7\textheight]{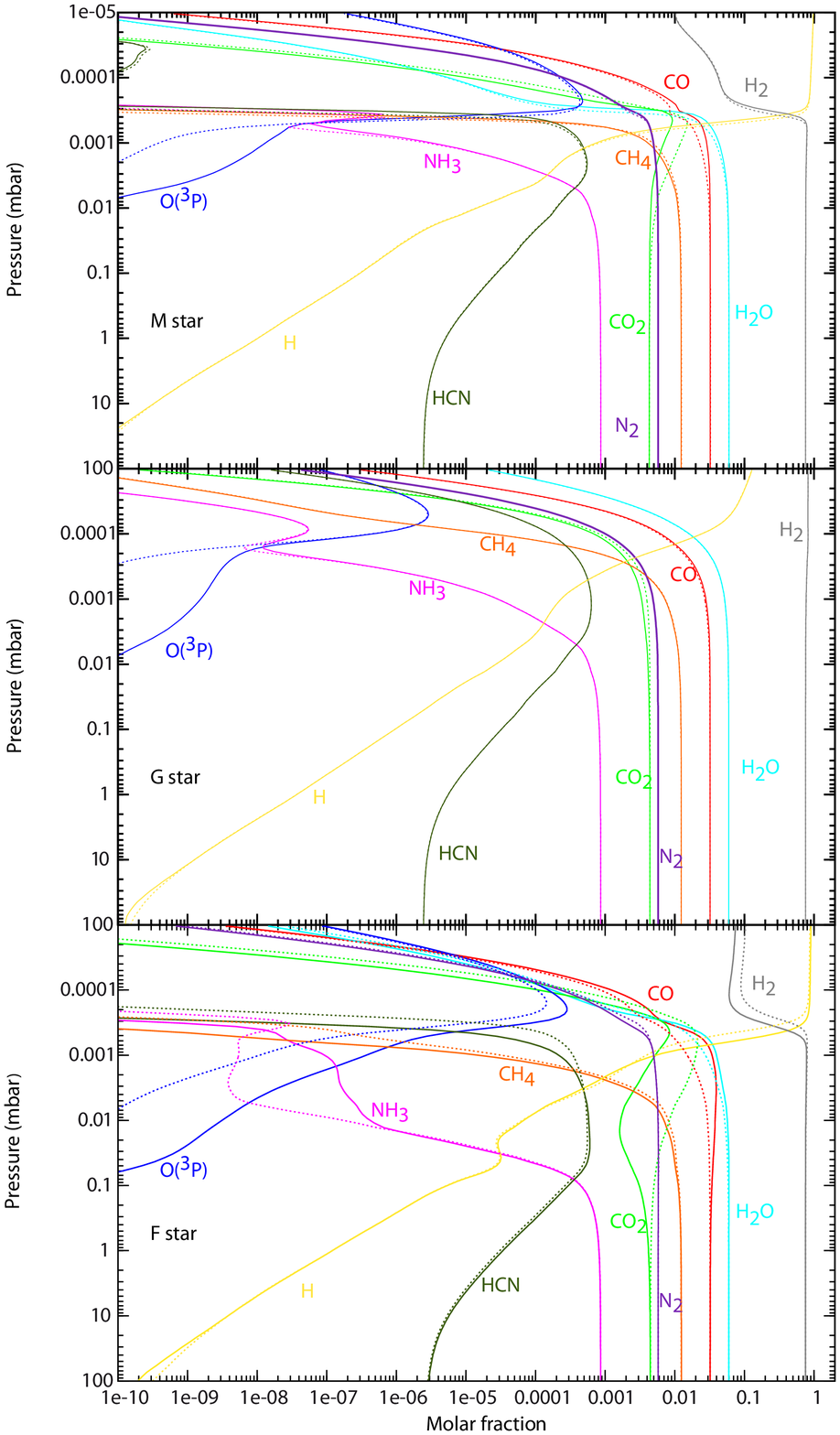}
\includegraphics[height=0.7\textheight]{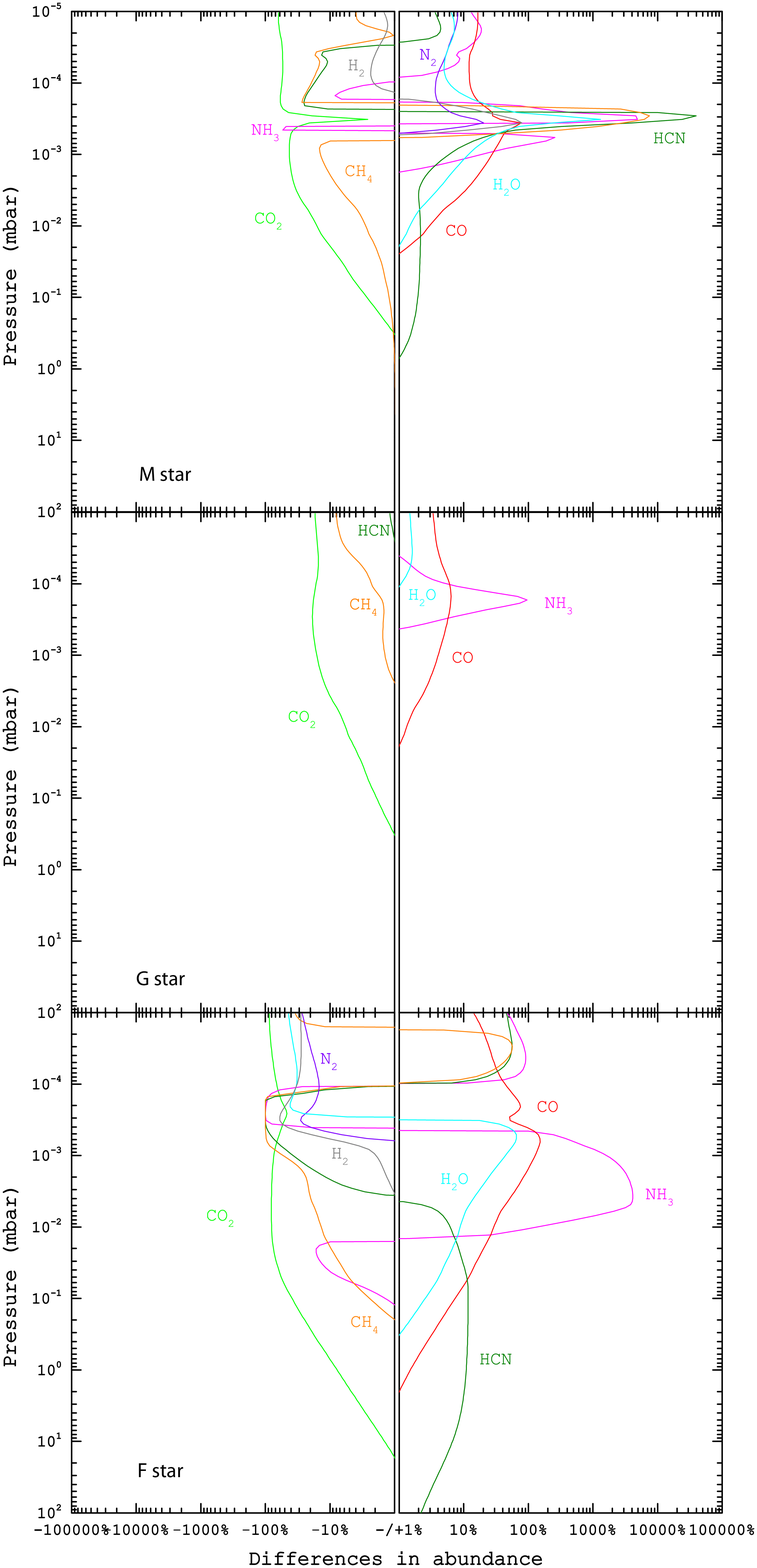}
\caption{\textit{Left:} Comparison of the atmospheric composition of GJ 436b using $\sigma_{\ce{CO2}}$(300~K) (dotted line) and $\sigma_{\ce{CO2}}$(550~K) (full line) when the planet orbits an M star (top), a G star (middle) and an F star (bottom). \textit{Right:} Differences in abundances (in \%) between the results obtained with $\sigma_{\ce{CO2}}$(300~K) and $\sigma_{\ce{CO2}}$(550~K) for species that have an abundance superior to $10^{-10}$, for an M star (top), a G star (middle), and an F star (bottom).}
\label{fig:GJ436b}
\end{figure*}

As we see in Fig.~\ref{fig:GJ436b} (left), changing the absorption cross section of \ce{CO2} has consequences on the abundances of some species. For instance, when considering an M star at $5\times10^{-4}$ mbar with $\sigma_{\ce{CO2}}$(300~K), \ce{CO2} has an abundance of $1.5 \times10^{-2}$, whereas with $\sigma_{\ce{CO2}}$(550~K), its abundance is only $0.82\times10^{-2}$, which corresponds to a reduction of 45\%. The compounds O($^3$P), CO, and \ce{NH3} are also affected by the change of $\sigma_{\ce{CO2}}(T)$ at the same pressure. There is almost no difference in the case of the G star. But for the F star at 0.01 mbar, \ce{CO2} is nearly eight times less abundant with $\sigma_{\ce{CO2}}$(550~K) than with $\sigma_{\ce{CO2}}$(300~K). Consequently, because \ce{CO2} absorbs more photons, less \ce{NH3} is destroyed and is 40 times more abundant than with $\sigma_{\ce{CO2}}$(300~K). We see that other species are also affected by the change of \ce{CO2} absorption cross section, such as \ce{CH4}, HCN, \ce{H2}, and \ce{N2}. To see these variations, we represented in Fig.~\ref{fig:GJ436b} (right) the differences of abundances of the major species (i.e., for species with an abundance superior to $10^{-10}$) between the two models. We see that the differences in abundances can reach almost $10^{5}$\%, even for species which are not directly linked to $\sigma_{\ce{CO2}}(T)$, such as HCN in the case of the M star.

\begin{figure}[!ht]
\includegraphics[width=1\columnwidth]{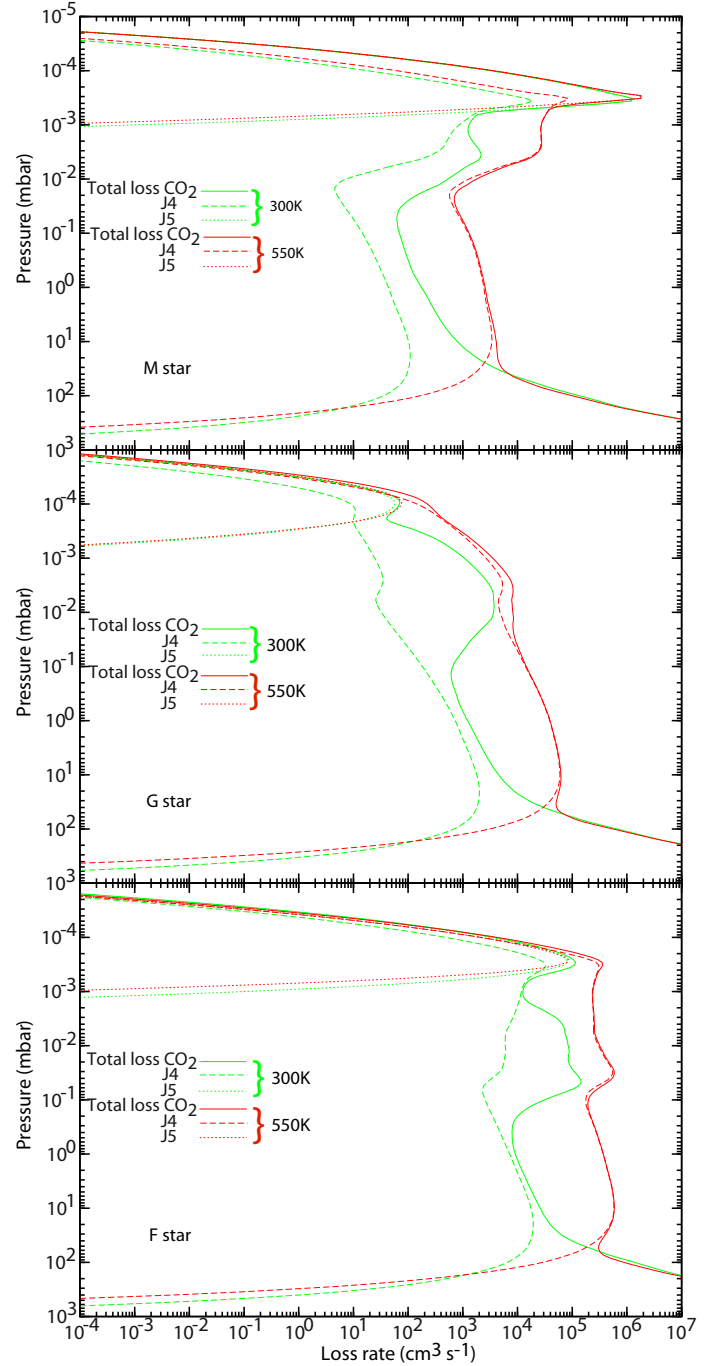}
\caption{Total loss rate and photodissociation rates of \ce{CO2} when using $\sigma_{\ce{CO2}}$(300~K) (green) and $\sigma_{\ce{CO2}}$(550~K) (red) in the atmosphere of GJ~436b when the planet orbits an M star (top), a G star (middle), and an F star (bottom).}
\label{fig:tauxperte}
\end{figure}

These differences are easily comprehensible. We chose the case of \ce{NH3} to illustrate it. At high temperature, the absorption cross section of \ce{CO2} is higher around 120~nm and for wavelengths superior to 150~nm, so \ce{CO2} absorbs more UV flux than with the ambient cross section. Consequently, more \ce{CO2} is photolysed. The UV photons that are now absorbed by \ce{CO2} were absorbed by \ce{NH3} when using $\sigma_{\ce{CO2}}$(300~K). Now \ce{NH3} absorbs fewer UV photons, so less is destroyed. This can be generalized to the other species. Indeed, the general behavior of these curves can be explained in terms of the various opacity sources that peak at slightly different altitudes. Because of the competition among the different opacity sources in the atmosphere, all the species absorbing in the same range of wavelength as \ce{CO2} are affected.\\
We see in Fig.~\ref{fig:tauxperte} that for pressures higher than $\sim$ 50~mbar, the total loss rate is not altered by the change of absorption cross section. For pressures lower than that, the total loss rate of \ce{CO2} is $\sim$ 10 times more important with $\sigma_{\ce{CO2}}(550~K)$ than with $\sigma_{\ce{CO2}}$(300~K). Between 10 and $5\times10^{-4}$ mbar, the loss rate of the photolysis process $J4$ is approximately equal to the total loss rate when we use $\sigma_{\ce{CO2}}$(550~K), whereas it is far from the case with $\sigma_{\ce{CO2}}$(300~K). Indeed, between 10$^{-3}$ and 10 mbar, the loss rate of $J4$ increases by a factor of 10-100 when we use $\sigma_{\ce{CO2}}$(550~K). At around $5\times 10^{-4}$ mbar with $\sigma_{\ce{CO2}}$(550~K),  the photodissociation process $J5$ is about 10 times less efficient in destroying \ce{CO2} than with $\sigma_{\ce{CO2}}$(300~K). Nevertheless, we do not see any change in the abundance of O($^1$D).
In both cases, with the M star we see a peak of destruction of \ce{CO2} around $3\times10^{-4}$ mbar. 

It is with the G star that the difference of composition is the least important when we change $\sigma_{\ce{CO2}}$. This is quite normal because the flux received by the planet before 200~nm is lower than with the M star and the F star (see Fig.~\ref{fig:GJ436b}). On the contrary, the difference of composition is the most important in the case F star, because it is the most important flux between 115 and 230 nm.

Since the VUV flux is more important with the F star than with the two previous cases, we observe in Fig.~\ref{fig:tauxperte} that the total loss rate of \ce{CO2} is approximately two times more important than for the G star case in the whole atmosphere. The total loss rate of \ce{CO2} in the F star case is roughly twice as high as than in the M star case in the entire atmosphere, except for pressures around 10$^{-3}$ mbar, where it is more than three times lower. As for the two other cases, we observe that with $\sigma_{\ce{CO2}}$(550~K), the photodissociation rate of $J4$ is much more important.


\section{Conclusion}

We measured absorption cross sections of carbon dioxide at high temperatures for the first time, in the range 115-200~nm with a resolution of 0.05~nm and in the range 195-230 nm with a resolution of 0.3~nm.
Between 115 and 200~nm, we measured $\sigma_{\ce{CO2}}(\lambda, T)$ at four temperatures: 300, 410, 480, and 550~K. For longer wavelengths, we made measurements at the following temperatures: 465, 510, 560, 610, 655, 750, and 800~K.  For $\lambda >$ 160~nm, we clearly see that the absorption cross section increases with the temperature. Thanks to the quasi-linear variation of $\ln(\sigma_{\ce{CO2}}(\lambda, T) \times \frac{1}{Q_v(T)})$ after 170~nm, we parametrize the variation of  $\ln(\sigma_{\ce{CO2}}(\lambda, T) \times \frac{1}{Q_v(T)})$ with a linear regression which allows us to calculate $\sigma_{\ce{CO2}}(\lambda, T)$ at any temperature in the range 170-230~nm.
As we show for GJ~436b, these new data have a considerable influence on the loss rate of \ce{CO2}, and on the atmospheric composition of exoplanets that possess high atmospheric temperatures. Placing a hot Neptune around different stars (M, G, and F), we find that  the F star is the star for which the change of absorption cross section has the most influence. This information is important to model other planets, like HD~221287b \citep{Naef2007}, HD~31253b \citep{meschiari2011lick}, or HD~153950b \citep{Moutou2009} which are orbiting F stars around 1.26 AU, so receive approximatively the same UV fluxes as in our simulations.\\

To model hot exoplanets, we recommend using cross sections relevant to the atmospheric temperature when available, or at least, as close as possible to the atmospheric temperature. Carbon dioxide is not the only absorbing species of exoplanet atmospheres. 
The influence of the absorption cross section of \ce{CO2} on the atmospheric composition of GJ~436b is only illustrative because the photochemistry results from the fact that species shield each other according to their abundances and their cross sections. We expect that the effect of $\sigma_{\ce{CO2}}(\lambda, T)$ will be more important on other types of atmospheres, in particular \ce{CO2}-rich atmospheres. But the real impact of the temperature dependence of $\sigma_{\ce{CO2}}(\lambda, T)$ can be evaluated only by taking into account the temperature dependence of all the other cross sections. Here, we simply show that it is necessary to establish this dependence for all species that absorb UV radiation. This work on \ce{CO2} is a first step towards this goal. Because \cite{venot2012} show that \ce{NH3} is an important absorber around 200 nm and that it absorbs UV flux very deep in the atmosphere (in pressure regions that can be probed with observations), we plan to measure the absorption cross section of this molecule at temperatures higher than 300~K. Finally, a great deal of work remains to be done in this area which is essential for the photochemical modeling of hot exoplanet atmospheres, whether terrestrial or gaseous.

\begin{acknowledgements} 
The authors wish to thank Gerd Reichard and Peter Baumg\"artel for their excellent assistance during the synchrotron radiation beam time periods. We acknowledge the financial support of the European Commission Programme "Access to Research Infrastructures" for providing access to the synchrotron facility BESSY in Berlin. We also acknowledge the financial support of the program PIR EPOV
and of the European Cooperation in Science and Technology - Chemistry and Molecular Sciences and Technologies (COST-CMST). O. V., F. S. and E. H. acknowledge support from the European Research Council (ERC Grant 209622: E$_3$ARTHs).

\end{acknowledgements}

\bibliographystyle{aa}
\bibliography{bib_article}

\begin{thebibliography}{39}
\expandafter\ifx\csname natexlab\endcsname\relax\def\natexlab#1{#1}\fi

\bibitem[{Beaulieu {et~al.}(2010)Beaulieu, Kipping, Batista, Tinetti, Ribas,
  Carey, Noriega-Crespo, Griffith, Campanella, Dong,
  {et~al.}}]{beaulieu2010water}
Beaulieu, J., Kipping, D., Batista, V., {et~al.} 2010, Monthly Notices of the
  Royal Astronomical Society, 409, 963

\bibitem[{{Beaulieu} {et~al.}(2011){Beaulieu}, {Tinetti}, {Kipping}, {Ribas},
  {Barber}, {Cho}, {Polichtchouk}, {Tennyson}, {Yurchenko}, {Griffith},
  {Batista}, {Waldmann}, {Miller}, {Carey}, {Mousis}, {Fossey}, \&
  {Aylward}}]{2011beaulieu}
{Beaulieu}, J.-P., {Tinetti}, G., {Kipping}, D.~M., {et~al.} 2011, The
  Astrophysical Journal, 731, 16

\bibitem[{Butler {et~al.}(2004)Butler, Vogt, Marcy, Fischer, Wright, Henry,
  Laughlin, \& Lissauer}]{butler2004neptune}
Butler, R., Vogt, S., Marcy, G., {et~al.} 2004, The Astrophysical Journal, 617,
  580

\bibitem[{Cossart-Magos {et~al.}(1992)Cossart-Magos, Launay, \&
  Parkin}]{cossart1992high}
Cossart-Magos, C., Launay, F., \& Parkin, J. 1992, Molecular Physics, 75, 835

\bibitem[{Cossart-Magos {et~al.}(2005)Cossart-Magos, Launay, \&
  Parkin}]{cossart2005high}
Cossart-Magos, C., Launay, F., \& Parkin, J. 2005, Molecular Physics, 103, 629

\bibitem[{Generalov {et~al.}(1963)Generalov, Losev, \&
  Maksimenko}]{generalov1963absorption}
Generalov, N., Losev, S., \& Maksimenko, V. 1963, Optics and Spectroscopy, 15,
  12

\bibitem[{Grevesse \& Sauval(1998)}]{grevesse1998standard}
Grevesse, N. \& Sauval, A. 1998, Space Science Reviews, 85, 161

\bibitem[{{Gueymard}(2004)}]{gueymard2004}
{Gueymard}, C. 2004, Solar Energy, 76, 423

\bibitem[{Hersant {et~al.}(2004)Hersant, Gautier, \&
  Lunine}]{hersant2004enrichment}
Hersant, F., Gautier, D., \& Lunine, J. 2004, Planetary and Space science, 52,
  623

\bibitem[{Huebner {et~al.}(1992)Huebner, Keady, \& Lyon}]{huebner1992solar}
Huebner, W., Keady, J., \& Lyon, S. 1992, Solar photo rates for planetary
  atmospheres and atmospheric pollutants (Kluwer Academic Pub)

\bibitem[{Jeffries {et~al.}(2005)Jeffries, Schulz, Mattison, Oehlschlaeger,
  Bessler, Lee, Davidson, \& Hanson}]{jeffries2005uv}
Jeffries, J., Schulz, C., Mattison, D., {et~al.} 2005, Proceedings of the
  Combustion Institute, 30, 1591

\bibitem[{Jensen {et~al.}(1997)Jensen, Guettler, \&
  Lyman}]{jensen1997ultraviolet}
Jensen, R., Guettler, R., \& Lyman, J. 1997, Chemical physics letters, 277, 356

\bibitem[{Koshi {et~al.}(1991)Koshi, Yoshimura, \& Matsui}]{Koshi1991519}
Koshi, M., Yoshimura, M., \& Matsui, H. 1991, Chemical Physics Letters, 176,
  519

\bibitem[{Lewis \& Carver(1983)}]{Lewis1983297}
Lewis, B. \& Carver, J. 1983, Journal of Quantitative Spectroscopy and
  Radiative Transfer, 30, 297

\bibitem[{Lewis {et~al.}(2010)Lewis, Showman, Fortney, Marley, Freedman, \&
  Lodders}]{lewis2010atmospheric}
Lewis, N., Showman, A., Fortney, J., {et~al.} 2010, The Astrophysical Journal,
  720, 344

\bibitem[{Line {et~al.}(2011)Line, Vasisht, Chen, Angerhausen, \&
  Yung}]{line2011thermochemical}
Line, M., Vasisht, G., Chen, P., Angerhausen, D., \& Yung, Y. 2011, The
  Astrophysical Journal, 738, 32

\bibitem[{Madhusudhan \& Seager(2011)}]{madhusudhan2011highGJ436b}
Madhusudhan, N. \& Seager, S. 2011, The Astrophysical Journal, 729, 41

\bibitem[{Meschiari {et~al.}(2011)Meschiari, Laughlin, Vogt, Butler, Rivera,
  Haghighipour, \& Jalowiczor}]{meschiari2011lick}
Meschiari, S., Laughlin, G., Vogt, S., {et~al.} 2011, The Astrophysical
  Journal, 727, 117

\bibitem[{Moses {et~al.}(2011)Moses, Visscher, Fortney, Showman, Lewis,
  Griffith, Klippenstein, Shabram, Friedson, Marley,
  {et~al.}}]{moses2011disequilibrium}
Moses, J., Visscher, C., Fortney, J., {et~al.} 2011, The Astrophysical Journal,
  737, 15

\bibitem[{{Moutou} {et~al.}(2009){Moutou}, {Mayor}, {Lo Curto}, {Udry},
  {Bouchy}, {Benz}, {Lovis}, {Naef}, {Pepe}, {Queloz}, \&
  {Santos}}]{Moutou2009}
{Moutou}, C., {Mayor}, M., {Lo Curto}, G., {et~al.} 2009, Astronomy \&
  Astrophysics, 496, 513

\bibitem[{{Naef} {et~al.}(2007){Naef}, {Mayor}, {Benz}, {Bouchy}, {Lo Curto},
  {Lovis}, {Moutou}, {Pepe}, {Queloz}, {Santos}, \& {Udry}}]{Naef2007}
{Naef}, D., {Mayor}, M., {Benz}, W., {et~al.} 2007, Astronomy \& Astrophysics,
  470, 721

\bibitem[{Oehlschlaeger {et~al.}(2004)Oehlschlaeger, Davidson, Jeffries, \&
  Hanson}]{oehlschlaeger2004ultraviolet}
Oehlschlaeger, M., Davidson, D., Jeffries, J., \& Hanson, R. 2004, Chemical
  physics letters, 399, 490

\bibitem[{Parkinson {et~al.}(2003)Parkinson, Rufus, \& Yoshino}]{parkinson2003}
Parkinson, W.~H., Rufus, J., \& Yoshino, K. 2003, Chemical Physics, 290, 251

\bibitem[{Reichardt {et~al.}(2001)Reichardt, Noll, Packe, Rotter, Schmidt, \&
  Gudat}]{reichardt2001b}
Reichardt, G., Noll, T., Packe, I., {et~al.} 2001, Nucl. Instrum. Methods Phys.
  Res. Sect. A 467-468, 458

\bibitem[{Schulz {et~al.}(2002)Schulz, Koch, Davidson, Jeffries, \&
  Hanson}]{Schulz200282}
Schulz, C., Koch, J., Davidson, D., Jeffries, J., \& Hanson, R. 2002, Chemical
  Physics Letters, 355, 82

\bibitem[{Segura {et~al.}(2005)Segura, Kasting, Meadows, Cohen, Scalo, Crisp,
  Butler, \& Tinetti}]{segura2005biosignatures}
Segura, A., Kasting, J., Meadows, V., {et~al.} 2005, Astrobiology, 5, 706

\bibitem[{Segura {et~al.}(2003)Segura, Krelove, Kasting, Sommerlatt, Meadows,
  Crisp, Cohen, \& Mlawer}]{segura2003ozone}
Segura, A., Krelove, K., Kasting, J., {et~al.} 2003, Astrobiology, 3, 689

\bibitem[{Southworth(2010)}]{southworth2010homogeneous}
Southworth, J. 2010, Monthly Notices of the Royal Astronomical Society

\bibitem[{Stark {et~al.}(2007)Stark, Yoshino, Smith, \& Ito}]{stark2007}
Stark, G., Yoshino, K., Smith, P.~L., \& Ito, K. 2007, Journal of Quantitative
  Spectroscopy \& Radiative Transfer, 103, 67

\bibitem[{Stevenson {et~al.}(2012)Stevenson, Harrington, Lust, Lewis,
  Montagnier, Moses, Visscher, Blecic, Hardy, Cubillos,
  {et~al.}}]{stevenson2012two}
Stevenson, K., Harrington, J., Lust, N., {et~al.} 2012, The Astrophysical
  Journal, 755, 9

\bibitem[{Stevenson {et~al.}(2010)Stevenson, Harrington, Nymeyer, Madhusudhan,
  Seager, Bowman, Hardy, Deming, Rauscher, \& Lust}]{stevenson2010possible}
Stevenson, K., Harrington, J., Nymeyer, S., {et~al.} 2010, Nature, 464, 1161

\bibitem[{Swain {et~al.}(2009{\natexlab{a}})Swain, Tinetti, Vasisht, Deroo,
  Griffith, Bouwman, Chen, Yung, Burrows, Brown, {et~al.}}]{swain2009water}
Swain, M., Tinetti, G., Vasisht, G., {et~al.} 2009{\natexlab{a}}, The
  Astrophysical Journal, 704, 1616

\bibitem[{Swain {et~al.}(2008)Swain, Vasisht, \& Tinetti}]{swain2008presence}
Swain, M., Vasisht, G., \& Tinetti, G. 2008, Nature, 452, 329

\bibitem[{Swain {et~al.}(2009{\natexlab{b}})Swain, Vasisht, Tinetti, Bouwman,
  Chen, Yung, Deming, \& Deroo}]{swain2009molecular}
Swain, M., Vasisht, G., Tinetti, G., {et~al.} 2009{\natexlab{b}}, The
  Astrophysical Journal Letters, 690, L114

\bibitem[{Tinetti {et~al.}(2010)Tinetti, Deroo, Swain, Griffith, Vasisht,
  Brown, Burke, \& McCullough}]{Tinetti2010}
Tinetti, G., Deroo, P., Swain, M.~R., {et~al.} 2010, The Astrophysical Journal
  Letters, 712, L139

\bibitem[{Tinetti {et~al.}(2007{\natexlab{a}})Tinetti, Liang, Vidal-Madjar,
  Ehrenreich, Lecavelier~des Etangs, \& Yung}]{tinetti2007infrared}
Tinetti, G., Liang, M., Vidal-Madjar, A., {et~al.} 2007{\natexlab{a}}, The
  Astrophysical Journal Letters, 654, L99

\bibitem[{Tinetti {et~al.}(2007{\natexlab{b}})Tinetti, Vidal-Madjar, Liang,
  Beaulieu, Yung, Carey, Barber, Tennyson, Ribas, Allard,
  {et~al.}}]{tinetti2007water}
Tinetti, G., Vidal-Madjar, A., Liang, M., {et~al.} 2007{\natexlab{b}}, Nature,
  448, 169

\bibitem[{Venot {et~al.}(2012)Venot, H\'ebrard, Agundez, Dobrijevic, Selsis,
  Hersant, \& Bounaceur}]{venot2012}
Venot, O., H\'ebrard, E., Agundez, M., {et~al.} 2012, Astronomy \&
  Astrophysics, 546, A43

\bibitem[{Yoshino {et~al.}(1996)Yoshino, Esmond, Sun, Parkinson, Ito, \&
  Matsui}]{yoshino96b}
Yoshino, K., Esmond, J.~R., Sun, Y., {et~al.} 1996, J. Quant. Spectrosc.
  Radiat. Transfer, 55, 53

\end{thebibliography}

\end{document}